\documentclass[a4paper,11pt]{article}
\usepackage{pos}

\title{Lattice calculation of electromagnetic corrections to $K\ell3$ decay}
\ShortTitle{Electromagnetic corrections to $K_{\ell3}$ decay}

\author*[a]{Norman H. Christ}
\author[b]{Xu Feng}
\author[c]{Luchang Jin}
\author[d]{Christopher T. Sachrajda}
\author[e]{Tianle Wang}

\affiliation[a]{Department of Physics, Columbia University 
 New York, New York 10027, USA}

\affiliation[b]{School of Physics, Peking University, Beijing 100871, China \\
Collaborative Innovation Center of Quantum Matter, Beijing 100871, China \\
Center for High Energy Physics, Peking University, Beijing 100871, China}

\affiliation[c]{Department of Physics, University of Connecticut, Storrs, Connecticut 06269, USA}
 
 \affiliation[d]{Department of Physics and Astronomy, University of Southampton,
Southampton SO17 1BJ, United Kingdom}

\affiliation[e]{Computational Science Initiative, Brookhaven National Laboratory, Upton, New York 11973, USA}

\emailAdd{nhc@phys.columbia.edu}
\emailAdd{xu.feng@pku.edu.cn}
\emailAdd{ljin.luchang@gmail.com}
\emailAdd{cts@phys.soton.ac.uk}
\emailAdd{twang3@bnl.gov}

\abstract{We describe a first-principles method to apply lattice QCD to compute the 
order $\alpha_{\mathrm{EM}}$ corrections to $K\to\pi\ell\nu_\ell$ decay.  
This method formulates the calculation in infinite volume with the conventional 
infinite-volume, continuum treatment of QED.  Infinite volume reconstruction 
is used to replace the QCD components of the calculation with finite-volume
amplitudes which can be computed in Euclidean space using lattice QCD,
introducing finite-volume errors which vanish exponentially as the volume 
used in the QCD calculation is increased.  This approach has also been 
described in an appendix to the recent paper: Phys.Rev.D 108 (2023) 1, 014501.}

\FullConference{The 40th International Symposium on Lattice Field Theory (Lattice 2023)\\
July 31st - August 4th, 2023\\
Fermi National Accelerator Laboratory\\}

\begin{document}
\maketitle

\section{Combining QCD and electromagnetism}

Given the increasing precision with which many important quantities can be computed using lattice QCD there is strong motivation to extend these calculations to include the order $\alpha_{\mathrm{EM}}$ effects of electromagnetism (E\&M).  While it is natural to apply the methods of lattice gauge theory to the combined $SU(3)\times U(1)$ strong and electromagnetic gauge group, the massless photon creates substantial obstacles to such a naive approach.  The combination of the preferred, translationally-invariant finite-volume periodic boundary conditions with Coulomb's law requires that only electrically neutral systems be studied.  This limitation can be avoided by using the QED$_{\mathrm L}$~\cite{Hayakawa:2008an} formulation of QED where one drops the $\vec k=0$ mode from the photon degrees of freedom at the cost of introducing finite-volume corrections which decrease only as inverse powers of the lattice volume -- power-law corrections which alter the Coulomb potential even at short distance adding $c_0/L$ and $c_2 r^2/L^3$ terms to the usual $1/r$ short-distance behavior, where $c_0$ and $c_2$ are constants.  For recent discussion of these $1/L^3$

In this talk we demonstrate an alternative strategy, referred to as QED$_\infty$ in which QCD and QED are combined but treated very differently~\cite{Bernecker:2011gh, Asmussen:2016lse, Blum:2017cer,RBC:2018dos}.  In this approach one begins by considering a entirely physical, Minkowski-space calculation performed in infinite volume with a physical, continuum formulation of QED.  (Since we intend to use perturbation theory to expand in $\alpha_{\mathrm{EM}}$ there is no need to treat QED non-perturbatively.)  We separate the physical Minkowski-space amplitude into QCD and QED factors and and attempt to express the QCD factor as an amplitude which can be computed in Euclidean space.  Such a transformation to Euclidean space may be accomplished through a Wick rotation in which Cauchy's theorem is used to show the identity of the product of QED and QCD amplitudes before and after rotation.  This treatment may even be possible when this Wick rotation is prevented by the contribution of single-particle states if they can be explicitly subtracted and their correct Minkowski-space contribution calculated separately.

The resulting combined Minkowski- and Euclidean-space calculation is still formulated in infinite volume without approximation.  In the next step one examines the Euclidean-space calculation of the QCD factor and explores whether the mass gap in QCD provides sufficient convergence as the vertices in the QCD amplitude are separated that the entire QCD factor can be computed in a finite volume with only exponentially suppressed finite-volume corrections.  If this is the case then we have the best of both worlds:  a finite-volume, Euclidean space calculation of the QCD factor that can be performed using lattice QCD and an infinite-volume QED calculation that can be performed in the continuum using standard Feynman rules for the photon and any lepton propagators.   

\begin{figure}[h]
\centering
\includegraphics[width=0.5\textwidth]{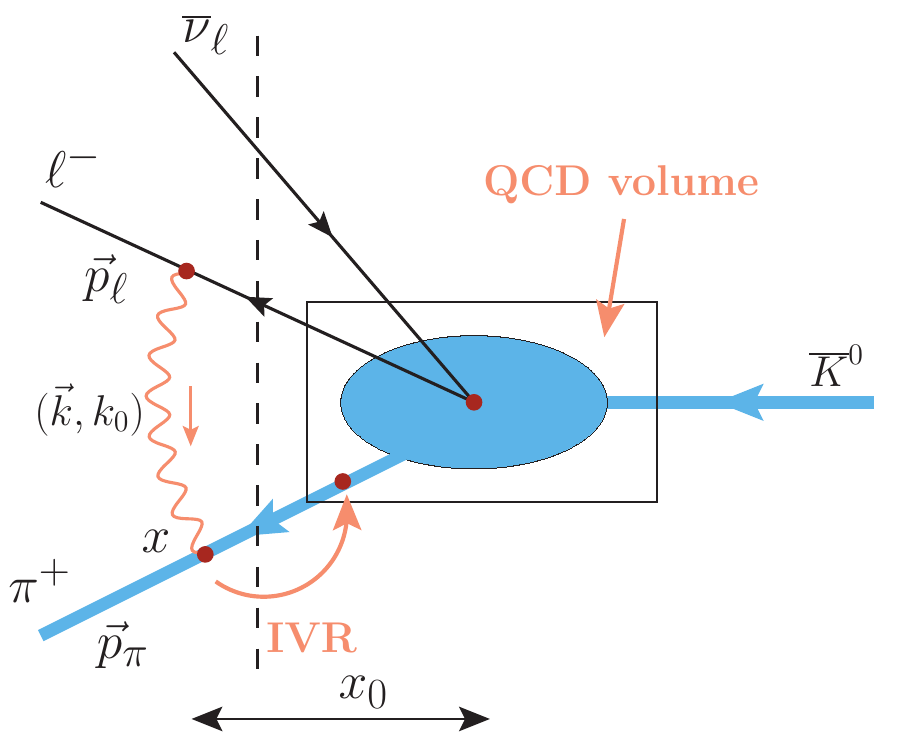}
\caption{A Feynman diagram showing the problematic photon-exchange topology in which the photon is exchanged between the lepton and a quark.  The momentum routing shown corresponds to that used in Eq.~\eqref{eq:kl3-IA}.  The vertical dotted line cuts through the three particle which complicate the Euclidean-space treatment of this process when the sum of their energies is near or below the mass of the kaon.}
\label{fig:feynman}
\end{figure}

Figure~\ref{fig:feynman} suggests this QCD $\times$ QED factorization for the E\&M corrections to $K\to\pi\ell\nu_\ell$ decay which is the topic of this paper.  The finite rectangular ``QCD volume'' suggests the limited region in which the Euclidean-space lattice QCD calculation will be performed while the lepton and photon propagators are treated in infinite-volume Minkowski space with neither finite-volume nor finite lattice spacing errors.  The weak vertex (the point from which the lepton and neutrino lines emerge) is intentionally placed at the center of the QCD volume to minimize the contribution of the space-time region where a hadronic E\&M current approaches the boundary of that volume and finite-volume errors arise because of unphysical periodic behavior of the finite-volume pion propagator.

The coupling shown in the figure between the photon and the emitted pion at the position labeled $x$ creates an important problem that naively would make QED$_\infty$ inapplicable to this process.  While there is exponential suppression associated with the propagation of the pion from the location of the weak vertex to the location of the interpolating operator which absorbs the final pion, that exponential suppression is present in the amplitude without regard to the location of the point on the pion propagator where the photon is absorbed.   As that photon-pion interaction point moves away from the weak vertex the resulting amplitude is only mildly suppressed by the power-law fall-off of the propagating photon.  However, at such large times the pion-photon vertex can move far from the weak vertex in space as well, with the resulting photon-pion interaction corrupted by the power-law finite-volume-distortion of the pion propagator as it moves close to the boundary of the finite, periodic QCD volume.  Interpreted in this way, the spatial extent of the QCD volume must be far larger than the time extent of the QCD volume if power-law-suppressed finite-volume corrections are to be avoided --- a potentially impractical requirement.  

This is the problem solved by infinite-volume reconstruction (IVR)~\cite{Feng:2018qpx}.  As is demonstrated below, the IVR technique explicitly limits the time extent of the QCD volume over which the pion-photon vertex is integrated.  The upper limit on the time is conventionally labeled $t_s$.   This restriction of the time integration is indicated by the arrow labeled IVR in Fig.~\ref{fig:feynman}.  If the time $x_0$ of this interaction point is also required to be sufficiently far from the weak vertex that only the pion propagation shown in the figure is possible, then a simple spatial Fourier transformation performed on the position dependence of that interaction point will extract all the information needed for the numerical integration of that interaction point over an infinite space-time volume with $x_0 > t_s$ to be accurately performed, thereby reconstructing the needed infinite-volume amplitude.

\section{Electromagnetic corrections to $K/\pi\to\ell\nu_\ell$ and $K\to\pi\ell\nu_\ell$}

After this introduction to the methods that we propose to use for the lattice QCD calculation of the E\&M corrections to  $K\to\pi\ell\nu_\ell$ decay, we now briefly discuss the status of the E\&M corrections to $K/\pi\to\ell\nu_\ell$ decay and the difficulties inherent in the more complex lattice calculation of the E\&M corrections to $K\to\pi\ell\nu_\ell$ decay, difficulties which have prevented the formulation of an {\it ab initio} lattice calculation of these corrections until now.

\underline{$K/\pi\to\ell\nu_\ell$}  A method for the calculation of the E\&M corrections to this decay was first proposed using QED$_{\mathrm{L}}$ and the calculation carried out by the Rome123-Southampton collaboration~\cite{Carrasco:2015xwa, Lubicz:2016xro, Giusti:2017dwk, DiCarlo:2019thl}.  Following similar methods a second independent calculation was presented in Ref.~\cite{Boyle:2022lsi} which also used QED$_{\mathrm{L}}$ but with physical quark masses.  The leading finite-volume correction depend on the lattice extent $L$ as $1/L$, a correction which is universal and can be computed in free scalar electrodynamics.  Corrections falling with higher powers of $1/L$ are also present but are structure dependent and at present unknown.  The size of the $1/L^3$ corrections is an important open question~\cite{DiCarlo:2021apt, Boyle:2022lsi}.  See also recent work presented at this conference~\cite{DiCarlo:2023rlz, DiCarlo:xxx, Portelli:xxx}.

In Ref.~\cite{Christ:2023lcc} we proposed in detail a second approach to this calculation using QED$_{\infty}$ and heavily exploiting IVR.  This approach offers two interesting advantages: (i) the finite-volume errors fall exponentially with increasing lattice extent and  (ii) the calculation can be organized so the all infrared divergences appear in the analytic portion of the calculation and cancel exactly --- the lattice-determined amplitudes are each infrared finite.  A calculation using these methods and physical quark masses is currently underway.  

In Appendix C of Ref.~\cite{Christ:2023lcc} we also describe how this same approach can be applied to the more difficult problem of finding the E\&M corrections to $K\to\pi\ell\nu_\ell$.  That appendix is the subject of this talk.  Using lattice methods to calculate these corrections faces two significant obstacles, which are most severe for the case of 
$\overline{K}^0\to\pi^+\ell^-\overline{\nu}_\ell$ decay where the final lepton and pion are charged.  Both problems result from the E\&M rescattering of the final $\pi^+$ and $\ell^-$, corresponding to the Feynman diagram topology shown in Fig.~\ref{fig:feynman}.  The first difficulty arises when the intermediate $\pi^+\ell^-\overline{\nu}_\ell$ state that is present before the exchange of the photon between the $\pi^+$ and $\ell^-$ carries an energy less than $M_K$.  Under these circumstances the Wick rotation of the integration contour followed by the loop energy $k_0$ is obstructed by exponential growth as a rotation of that contour is attempted.  (Note, these difficulties are also present when the pion is neutral but are less important since in that case the pion-photon coupling vanishes as $k^2\to0$.)

The second, related difficulty comes from those $\pi^+\ell^-\overline{\nu}_\ell$ intermediate states whose energy is close to $M_K$.   Such nearly on-shell states can travel without exponential suppression to the edges of a finite QCD volume and introduce potentially large finite-volume corrections.  This can be easily seen if one compares an infinite-volume calculation of the complex $\pi^+\ell^-$ rescattering (described by a singular integral with an imaginary part coming from an energy-conserving $\pi^+\ell^-\overline{\nu}_\ell$ state and a real part given as a principal value) with the finite-volume result which would be a sum of discrete finite-volume states each corresponding to a non-zero, but possibly small, non-covariant energy denominator.  We will now describe how these difficulties can be overcome.

\section{Overview of the solution}

Both of the difficulties identified in the previous section appear when the photon is exchanged between the lepton and a quark and when the quark-photon vertex, identified as $x$ in Fig.~\ref{fig:feynman}, is located at a time much later than the time of the weak vertex, taken here to be at the origin:  $x_0 \gg 0$.   In this problematic region, the dominant contribution will come from the single-pion intermediate state, allowing the QCD factor in the decay amplitude to be written:
\begin{eqnarray}
\mathcal{A}^{\mu\nu}_\pi(\vec p_\pi,\vec x, x_0) = \langle\pi(\vec p_\pi)|J^\mu_{\mathrm{EM}}(\vec x,x_0)
                            \left[\int d^3 p|\pi(\vec p)\rangle\langle\pi(\vec p)|\right]J^\nu_{\mathrm{W}}(0)|K(\vec 0)\rangle.
                            \label{eq:pion_contrib}
\end{eqnarray}
If the pion contribution identified in Eq.~\eqref{eq:pion_contrib} is explicitly subtracted from the Minkowski-space decay amplitude, then the subtracted amplitude can be Wick rotated without difficulty and the resulting Euclidean-space QCD amplitude will be localized with contributions from the region where the weak and E\&M vertices are separated by the space-time interval  $x$, falling exponentially as $\exp(-|x| E)$ in the Euclidean distance $|x|$ with the energy $E$ likely given by the $\rho$ mass. 

Thus, neither of the difficulties outlined above will appear in that portion of the calculation from which the pion contribution given in Eq.~\eqref{eq:pion_contrib} has been subtracted.  The subtracted amplitude can be correctly computed in Euclidean space and that Euclidean-space amplitude is exponentially localized so that any finite-volume errors will also be exponentially suppressed.  In this argument we do ignore the contribution of possible low-energy $\pi\pi\ell\overline{\nu}_\ell$ intermediate states.   These four-particle intermediate states are expected to be suppressed by 4-body phase space and to contribute at the few-percent level.  (The effects of a possible low-energy $\pi\pi\pi\ell\overline{\nu}_\ell$ intermediate state will be even smaller.)  In an eventual lattice calculations such contributions must be estimated phenomenologically and their omission included as a systematic error or they can be avoided altogether by restricting the final state kinematics to a region where one or both states cannot contribute.

As we discuss in the next section, the Minkowski-space amplitude given in Eq.~\eqref{eq:pion_contrib} can be computed from lattice QCD using IVR.  More specifically two calculations are required.  First this single-pion amplitude can be subtracted from the complete decay amplitude and their difference Wick-rotated and evaluated using lattice methods in Euclidean space.  Second, lattice QCD can be used to determine this single-pion amplitude in Minkowski space, compensating for this subtraction.  This explicit Minkowski-space contribution then provides a first-principles, non-perturbative result that correctly contains the complex pion-lepton rescattering effects with a real part coming from the proper infinite-volume principal part computation and an imaginary part arising from the usual energy-conserving delta function.  This dual-use of the IVR method allows both a conventional Euclidean lattice calculation of the short distance part of the decay amplitude with exponentially suppressed finite-volume errors and the Minkowski-space calculation of the pion-lepton scattering, including the non-perturbative effects of the pion form factor.

\section{Infinite-volume reconstruction of the single-pion contribution}

The IVR technique exploits the known space-time dependence of amplitudes which are dominated by a single-pion intermediate state.   Specifically we can express the critical, single-pion Minkowski amplitude given in Eq.~\eqref{eq:pion_contrib} as follows:
\begin{eqnarray}
\mathcal{A}^{\mu\nu}_\pi(\vec p_\pi,\vec x, x_0) &=& \langle\pi(\vec p_\pi)|J^\mu_{\mathrm{EM}}(\vec x,x_0)
                            \left[\int d^3 p|\pi(\vec p)\rangle\langle\pi(\vec p)|\right]J^\nu_{\mathrm{W}}(0)|K(\vec 0)\rangle_M.
                            \label{eq:pion_contrib-B} \\
                            && \hskip -0.6 in = \int d^3 p e^{-i(x_0+it_s)(E_{\vec p}-E_\pi)}
                            \langle\pi(\vec p_\pi)|J^\mu_{\mathrm{EM}}(\vec x, -it_s)|\pi(\vec p)\rangle
                            \langle\pi(\vec p)|J_{\mathrm{W}}^\nu(0)|K\vec 0\rangle_M \label{eq:pion_contrib-C} \\
                            && \hskip -0.6 in = \int d^3 p e^{-i(x_0+it_s)(E_{\vec p}-E_\pi)} \int \frac{d^3y}{(2\pi)^3}
                                                           e^{i(\vec p-\vec p_\pi)\cdot(\vec x - \vec y)} \label{eq:pion_contrib-D} \\
                            && \hskip 1.6 in h^{\mu\rho} h^{\nu\sigma}\langle\pi(\vec p_\pi)|J_{\mathrm{EM}}^\rho(\vec y, t_s)
                            J_{\mathrm{W}}^\sigma(0)|K\vec 0\rangle_E, \nonumber
\end{eqnarray}
where $E_{\vec p} = \sqrt{\vec p\,^2+M_\pi^2}$ and for simplicity we use $E_\pi = E_{\vec p_\pi}$.   The subscripts $M$ and $E$ identify QCD amplitudes that are computed in Minkowski and Euclidean space, respectively.  The tensor $h^{\alpha\beta}$ is a diagonal matrix with either $1$ or $i$ on the diagonal as needed to convert the Euclidean conventions for the current components in the Euclidean-space amplitude on the second line of Eq.~\eqref{eq:pion_contrib-D} into the Minkowski-space conventions used in the Minkowski-space amplitude $\mathcal{A}^{\mu\nu}_\pi$ which appears on the left-hand side of Eq.~\eqref{eq:pion_contrib-B}.  

Equation~\eqref{eq:pion_contrib-C} follows from Eq.~\eqref{eq:pion_contrib-B} by simply inserting $+it_s$ into the argument of the exponent and a canceling shift of $-it_s$ in the argument of $J_{\mathrm{EM}}$.  Equation~\eqref{eq:pion_contrib-D} is obtained from Eq.~\eqref{eq:pion_contrib-C} by recognizing the Minkowski-space amplitude with an imaginary time argument as actually a Euclidean amplitude and replacing the explicit insertion of a single-pion intermediate state carrying momentum $\vec p$ by a Fourier transform which projects onto the same state.  Of course, we must require that the time $t_s$ is sufficiently large that only single-pion intermediate states can contribute.

Equation~\eqref{eq:pion_contrib-D} provides the IVR result which we need.  The Minkowski-space single-pion contribution to the QCD matrix element for $x_0>0$ is expressed as the Fourier transform of a Euclidean amplitude that can be directly evaluated in a finite-volume lattice QCD calculation.  This result can be inserted into the Feynman amplitude represented by Fig.~\ref{fig:feynman} to give the contribution of this single-pion intermediate state to the E\&M correction to the $K\to\pi\ell\nu_\ell$ decay in which the photon is exchanged between the charged lepton and a quark:
\begin{eqnarray}
\mathcal{A}_{K\ell3}^I(\vec p_\pi, \vec p_\ell) &=& 
 \int\!\! d^4 k\, \int\!\! d^4 x \, \theta(x_0)  e^{-i(\vec x \cdot \vec k - x_0k_0)}
                                             \mathcal{A}^{\mu\nu}_\pi(\vec p_\pi, \vec x, x_0)  \nonumber  \\
   && \hspace{0.6in} \cdot\frac{1}{k^2-i\epsilon}\frac{\bar{u}_\ell(\vec p_\ell) \gamma_\mu\bigl(\gamma\cdot(p_\ell+k) +m_\ell\bigr)\gamma_\nu(1-\gamma^5)v_{\overline{\nu}}(\vec p_{\bar{\nu}})}{(p_\ell+k)^2 +m_\ell^2 -i\epsilon}.
\label{eq:kl3-IA} \\
&\equiv& \int\!\! d^4 x\, L(\vec p_\ell, x)_{\mu\nu}^M \, \mathcal{A}^{\mu\nu}_\pi(\vec p_\pi, \vec x, x_0).
\label{eq:kl3-IB}
\end{eqnarray} 
where Eq.~\eqref{eq:kl3-IB} provides a definition for the analytic ``leptonic'' factor $L(\vec p_\ell, x)_{\mu\nu}^M$ which is defined in Minkowski-space.  We have labeled this single-pion contribution $\mathcal{A}_{K\ell3}^I$

These same quantities can be used to write an explicit formula for the remaining contribution to this problematic diagram in which the photon is exchanged between the lepton and a quark, which we label $\mathcal{A}_{K\ell3}^{II}$.  This is the contribution from which the single-pion amplitude has been subtracted, allowing a Wick rotation to Euclidean space and subsequent lattice QCD evaluation:
\begin{eqnarray}
\mathcal{A}_{K\ell3}^{II}(\vec p_\pi, \vec p_\ell) 
&=& \int\!\! d^4 x \, L(\vec p_\ell, x)_{\mu\nu}^E
            \Bigl[\langle\pi(\vec p_\pi)|J_{\mathrm{EM}}^\mu(x) J_{\mathrm{W}}^\nu(0)|K(\vec 0)\rangle_E \label{eq:kl3-II} \\
&&\hskip 2.0 in                      - h^{\mu\rho} h^{\nu\sigma}\mathcal{A}^{\rho\sigma}_\pi(\vec p_\pi, \vec x, -ix_0) \Bigr]. 
                     \nonumber
\end{eqnarray} 
Here the subtraction of the single-pion amplitude $\mathcal{A}^{\rho\sigma}(\vec p_\pi, \vec x, -ix_0)$ results in a subtracted amplitude which falls sufficiently rapidly that the usual Wick rotation to Euclidean space is well-defined.  In Eq.~\eqref{eq:kl3-II} we have introduced a Euclidean-space version, $L(\vec p_\ell, x)_{\mu\nu}^E$ of the leptonic factor but simply evaluated the Minkowski-space amplitude $\mathcal{A}^{\rho\sigma}_\pi(\vec p_\pi, \vec x, x_0)$ at imaginary time.

An interesting issue raised by the finite-volume matrix elements which appear in Eqs.~\eqref{eq:pion_contrib-D} and \eqref{eq:kl3-II} is the choice of the out-going pion momentum $\vec p_\pi$.   Since our approach is put forward as one in which all finite-volume corrections are exponentially suppressed in the linear extent of the QCD volume, we should expect that any pion momentum $\vec p_\pi$ will be accessible.  In fact, this is the case.  We must introduce a local pion interpolating operator $\pi(x)$ and use its Fourier transform $\int\!\!d^3 x \exp(-i\vec p_\pi \cdot \vec x)$  to absorb the final-state pion when evaluating these matrix elements for the general three-momentum $\vec p_\pi$.  

While the integration volume used to perform this Fourier transform must be limited to the QCD volume, the exponential localization present in the Green's functions that would be used when evaluating Eq.~\eqref{eq:pion_contrib-D} or \eqref{eq:kl3-II} should result in this unwanted truncation of the Fourier transform introducing only exponentially small errors.   Of course, it is likely that these matrix elements change slowly as the external momentum $\vec p_\pi$ varies so evaluation at a few allowed lattice momenta $\vec p_\pi = 2\pi \vec n/L$
where $\vec n$ is a three-tuple of integers, should be sufficient.

\section{Conclusion}

We have presented for the first time a strategy to use lattice QCD to evaluate the electromagnetic corrections to the decay $K\to\pi\ell\nu_\ell$~\cite{Christ:2023lcc}.  The approach described uses infinite volume reconstruction~\cite{Feng:2018qpx} and introduces finite-volume errors which decrease exponentially as the linear extent of the volume used in the lattice calculation grows.  We have focused on the diagram which makes this calculation difficult for Euclidean-space lattice methods, where a photon is exchanged between the  final-state lepton and a quark.  This amplitude, if naively evaluated in Euclidean space will be dominated by an unphysical contribution in which the kaon decays into a pion, lepton and lepton-neutrino with total energy below that of the kaon.  This three-particle state then propagates for an extended time before the final photon exchange produces the state with the final kinematics. In the method proposed here infinite-volume reconstruction is used to determine the QCD Green's function in which the hadronic intermediate state which propagates from the weak current to the E\&M current carried by the quarks is a single pion.  This Green's function can then be used both to remove the unwanted $\pi\ell\nu_\ell$ state described above and to calculate the complex, Minkowski-space E\&M rescattering between the final lepton and pion -- the other challenging component of this calculation.  

In this presentation we have focused on these two issues which are present in the particular diagram shown in Fig.~\ref{fig:feynman}.   We expect that the other aspects of this calculation will be straight-forward applications of the methods already developed in Ref.~\cite{Christ:2023lcc}.   The methods in that paper are now being applied in a physical calculation of the E\&M corrections to $K/\pi\to\ell\nu_\ell$ decay, already a very challenging calculation.  The more complicated $K\to\pi\ell\nu_\ell$ calculation discussed in these proceedings must wait at least until that first calculation is complete.

\section*{Acknowledgements}

N.H.C.~acknowledges support from U.S.~DOE Grant No.~DE-32SC0011941.  X.F.~was supported in part by NSFC of China under Grants No.~12125501, No.~12070131001, and No.~12141501, and by the National Key Research and Development Program of China under No.~2020YFA0406400.   L.J.~acknowledges support under U.S.~DOE Grant No.~DESC0010339 and U.S.~DOE Office of Science Early Career Award No.~DE-SC0021147.   C.T.S.~was partially supported by an Emeritus Fellowship from the Leverhulme Trust and by STFC (UK) Grant No.~ST/T000775/1.

\bibliographystyle{JHEP}

\bibliography{Kl3-EM-2023}

\end{document}